\documentclass[conference]{IEEEtran}
\IEEEoverridecommandlockouts

\newcommand{\method}{SCOP}

\usepackage{url}
\usepackage{cite}
\usepackage{amsmath}
\usepackage{amssymb}
\usepackage{amsfonts}
\usepackage{algorithmic}
\usepackage{graphicx}
\usepackage{textcomp}
\usepackage{xcolor}
\usepackage{bm}
\usepackage{threeparttable}
\usepackage{adjustbox}
\usepackage{multicol}
\usepackage{multirow}
\usepackage{graphicx}
\usepackage{subfigure}
\usepackage{booktabs}
\usepackage{balance}
\usepackage{caption}

\def\BibTeX{{\rm B\kern-.05em{\sc i\kern-.025em b}\kern-.08em
    T\kern-.1667em\lower.7ex\hbox{E}\kern-.125emX}}
\begin{document}

\title{SCOP: A Sequence-Structure Contrast-Aware Framework for Protein Function Prediction}

\author{
    \IEEEauthorblockN{Runze Ma\IEEEauthorrefmark{2}\textsuperscript{*},
    Chengxin He\IEEEauthorrefmark{3},
    Huiru Zheng\IEEEauthorrefmark{5},
    Xinye Wang\IEEEauthorrefmark{2},
    Haiying Wang\IEEEauthorrefmark{5},
    Yidan Zhang\IEEEauthorrefmark{2},
    Lei Duan\IEEEauthorrefmark{2}\textsuperscript{*}}
    \IEEEauthorblockA{\IEEEauthorrefmark{2} School of Computer Science, Sichuan University, Chengdu, China}
    \IEEEauthorblockA{\IEEEauthorrefmark{3} College of Biomedical Engineering, Sichuan University, Chengdu, China}

    \IEEEauthorblockA{\IEEEauthorrefmark{5} School of Computing, Ulster University, Belfast, United Kingdom}
    \IEEEauthorblockA{* Corresponding authors. Email: runze.ma@outlook.com, leiduan@scu.edu.cn}
}

\maketitle

\begin{abstract}
Improving the ability to predict protein function can potentially facilitate research in the fields of drug discovery and precision medicine. 
Technically, the properties of proteins are directly or indirectly reflected in their sequence and structure information, especially as the protein function is largely determined by its spatial properties. 
Existing approaches mostly focus on protein sequences or topological structures, while rarely exploiting the spatial properties and ignoring the relevance between sequence and structure information. 
Moreover, obtaining annotated data to improve protein function prediction is often time-consuming and costly. To this end, this work proposes a novel contrast-aware pre-training framework, called SCOP, for protein function prediction. 
We first design a simple yet effective encoder to integrate the protein topological and spatial features under the structure view. Then a convolutional neural network is utilized to learn the protein features under the sequence view. 
Finally, we pretrain SCOP by leveraging two types of auxiliary supervision to explore the relevance between these two views and thus extract informative representations to better predict protein function. 
Experimental results on four benchmark datasets and one self-built dataset demonstrate that SCOP provides more specific results, while using less pre-training data.
\end{abstract}


\section{Introduction}
Proteins play a crucial role in physiological activities such as immune response, cell signaling and adhesion~\cite{figlia2020metabolites}, being the major carriers of all life activities. The implications of understanding the functions of proteins can help in protein-protein interactions, drug design and precision medicine~\cite{lu2020recent}.

The key to predicting protein function lies in extracting the representation of the protein, which depends on the definition of the protein, also known as protein descriptors~\cite{ma2022morn}. Typically, proteins can be described in the following forms: (\romannumeral1) sequence descriptors, which utilize amino acid sequences, \textit{e.g.}, in FASTA format~\cite{rao2019evaluating}; (\romannumeral2) structure descriptors, which provide a comprehensive description of the secondary or higher-level structures of proteins, \textit{e.g.}, protein contact maps depicting the two-dimensional topological structures of proteins, ball-and-stick models and ribbon diagrams portraying the three-dimensional spatial structures of proteins~\cite{zheng2019deep,torrisi2020deep}. Therefore, the current prevailing protein function prediction algorithms primarily rely on the sequence or structural features of proteins, while being optimized for specific data formats to enhance accuracy in predictions.

Protein function prediction has been the subject of extensive research for several years. Early methods~\cite{cozzetto2013protein} used predefined rules to extract physiochemical or evolutionary features. Nowadays, there are several deep learning-based works trying to predict protein functions. 
Most methods~\cite{rao2019evaluating,Bepler2021LearningTP,Meier2021LanguageME} leverage language models to learn representations from protein sequences and then predict their functions, whereas others~\cite{Gligorijevi2021Structurebased,xia2021geometric} learn the structural features of proteins to predict their functions. Despite the considerable progress made by existing methods in predicting protein functions, the following limitations are yet need to be considered:
\begin{itemize}
    \item \emph{Difficulties in overcoming the scarcity of protein labels}. Available data on physicochemical properties and biological functions of proteins is scarce, as such information is usually obtained by wet-lab experiments, which are generally time and cost intensive. 
    \item \emph{Deficiencies in learning the structure features of proteins}. The structure of a protein determines a wide range of protein properties. Existing sequence-based models fail to consider the protein's structure information, and most structure-based models only consider the 2D topological structures of proteins, ignoring the spatial features of specific conformations in the 3D space, meaning the learned representation is likely to be incomplete. 
    \item \emph{Inability in exploiting the relevance between protein sequences and structures}. Sequence and structure descriptors profile a protein at different levels.  However, existing methods either learn protein representations from one perspective alone, or simply perform feature extractions for sequences and structures in isolation. Such approaches do not exploit the relevance and associations between protein sequences and structures, making the learned representations potentially incomprehensive.
\end{itemize}

In order to tackle the above issues, we propose a novel protein function prediction approach, called SCOP (short for \underline{s}equence-stru\underline{c}ture c\underline{o}ntrast-aware \underline{p}re-training). The characteristics of SCOP include: (\romannumeral1) it introduces a protein structure-based encoder to integrate the protein topological and spatial features; (\romannumeral2) it makes full use of supervision contained in protein sequences paired with structures to explore the relevance between these two views; (\romannumeral3) it proposes a contrast-aware pre-training framework which can learn protein representations without label information.

The contributions of our work are fourfold:
\begin{itemize}
    \item We propose a contrast-aware pre-training model, \method, for utilizing protein sequence and structure features to predict its functions while using less pre-training data.    
    \item We design a novel protein encoder to learn protein features under the structure view, which can fully incorporate the topological and spatial structure features. 
    \item We first introduce two supervision tasks for protein pre-training, explicitly exploiting relevance and associations between sequences and structures of proteins.
    \item We evaluate the performance of \method \space on four real-world datasets and one original dataset. Extensive experimental results demonstrate that \method \space is effective in protein function prediction and superior to baselines on benchmarks.
\end{itemize}

\section{Related Work}
\label{rel}

\subsection{Sequence-Based Protein Function Prediction}
Proteins can be stored in FASTA format, which is a text-based format used to describe the protein sequence information. Protein representation learning based on sequence approaches have been studied for decades. Inspired by the advancement of natural language processing (NLP), some models (e.g., CNN, Transformer) have been employed in learning protein representations. Shanehsazzadeh \textit{et al}.~\cite{Shanehsazzadeh2020IsTL} developed a model for learning representations of proteins by a shallow convolutional neural network; Rao \textit{et al}.~\cite{rao2019evaluating} leveraged transformer models to learn protein representations and predict their structures and functional properties; Brandes \textit{et al}.~\cite{Brandes2021ProteinBERTAU} referred to the idea of BERT to predict the biophysical attributes of proteins.
    
\subsection{Structure-Based Protein Function Prediction}
The protein's structure determines its function, hence it is essential to learn protein representations by integrating the structural information. Derevyanko \textit{et al}.~\cite{Derevyanko2018DeepCN} tried to encode protein structure information by 3D CNNs. As proteins are able to represented as graphs, several approaches ~\cite{Baldassarre2020GraphQA, Gligorijevi2021Structurebased,Hermosilla2021IntrinsicExtrinsic,Zhang2022Protein} have been proposed to leverage graph neural networks to learn protein representations. Among these approaches, Hermosilla \textit{et al}.~\cite{Hermosilla2021IntrinsicExtrinsic} designed a graph neural network to incorporate both intrinsic and extrinsic distances among atoms in a protein. Zhang \textit{et al}.~\cite{Zhang2022Protein} transformed the atom-level protein graph to residue-level protein graph and applied a relational graph convolutional layer to learn protein representations. Another category of approaches~\cite{Sverrisson2020Fast} leveraged the protein surface information to extract its structure features.

\subsection{Multi-view Based Protein Function Prediction}
Protein descriptors contain comprehensive information, suggesting that multi-view learning can be beneficial for protein function prediction. Gligorijevic \textit{et al}.~\cite{Gligorijevi2021Structurebased} utilized a graph convolutional network that integrated protein structure and pre-trained sequence embeddings. Bepler \textit{et al}.~\cite{Bepler2021LearningTP} improved pre-trained sequence-based models by leveraging protein structural information during the pre-training period. Wang \textit{et al}.~\cite{Wang2022LMGVPAE} and Boadu \textit{et al}.~\cite{boadu2023combining} incorporated inductive biases from 3D structures of proteins during fine-tuning stage to combine a graph neural network with a protein language model.

\section{Methodology} 
\label{methods}

\subsection{Sequence View Guided Representation}
\label{subsec:seq}
Proteins can be saved in the FASTA format, serving as a type of protein sequence descriptor. We utilize FASTA strings as sequence features and apply a convolutional neural network to obtain protein representations under the sequence view. For a FASTA string $S$, an embedding layer $\phi(\cdot)$ is first applied to transform it into a vector. Then a convolutional neural network, denoted as $g_{seq}$ in Figure~\ref{fig:architecture}(a), is utilized to extract contextual information within the protein sequence. The expression can be depicted as:
\begin{align}
    \bm{h}^{S} = g_{seq}(\phi(S))
\end{align}
Here we can get the protein representation $\bm{h}^{S}$ under the sequence view, and it will be used for the pre-training module.

\subsection{Structure View Guided Representation}
\label{subsec:struct}
We introduce the concepts of the Protein Topological Graph (PTG) and Protein Spatial Graph (PSG) in the appendix\footnote{\url{https://github.com/mrzzmrzz/SCOP/blob/main/Appendix.pdf}}, describing protein structures from topological and spatial perspectives. As for a protein, we can construct its PTG, denoted as $\mathcal{G}$, and PSG, denoted as $\mathcal{M}$, to learn the topological and spatial features of the protein, respectively.

\begin{figure*}[!t]
    \centering
    \includegraphics[scale=0.24]{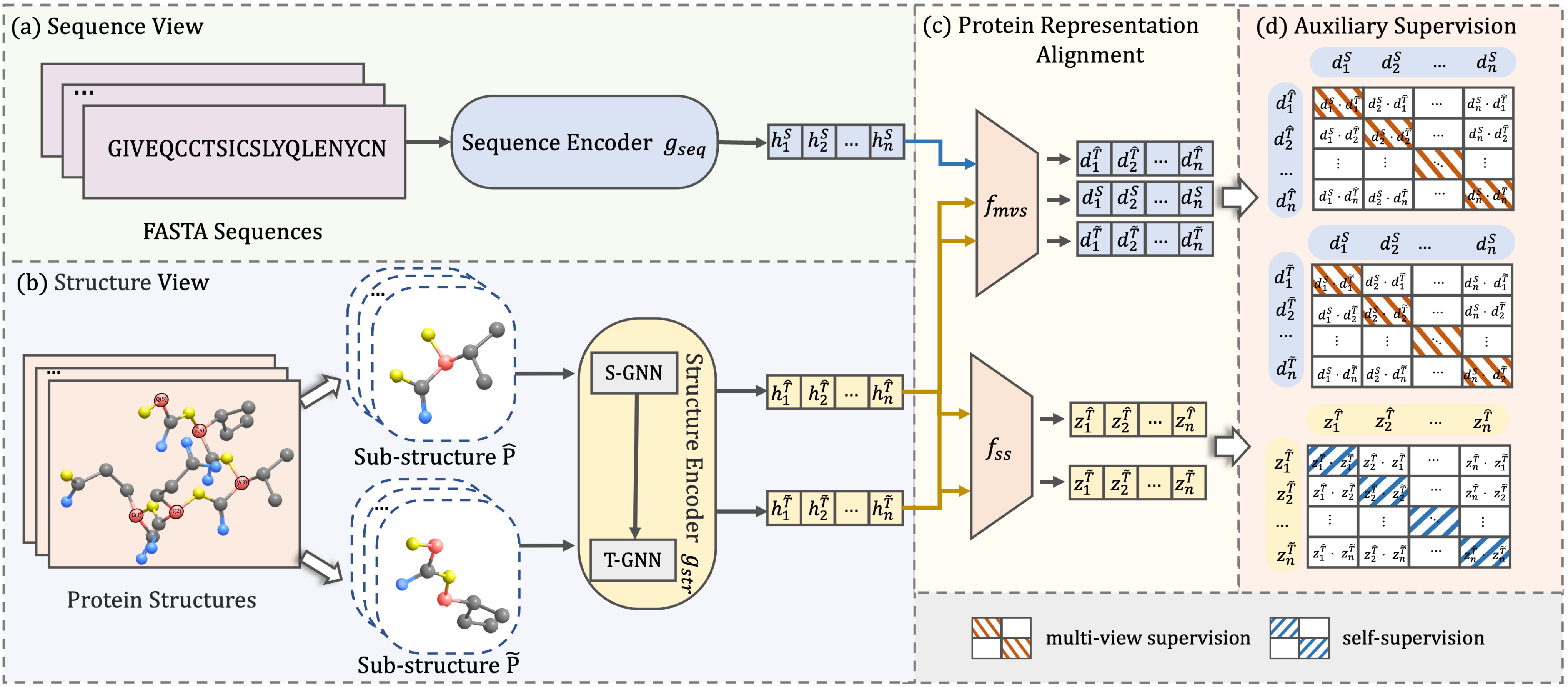}
    \caption{The proposed \method \space framework. It is composed of four parts: (a) learns the protein representation via sequence encoder $g_{seq}$. (b) adopts a stochastic augmentation module to sample two different sub-proteins $\hat{\mathcal{P}}$ and $\tilde{\mathcal{P}}$ then learns their representations by the structure encoder $g_{str}$. (c) adaptively projects the protein representation into a common embedding space. (d) conducts the self-supervision task under the structure view and multi-supervision task under the sequence-structure view.}
    \label{fig:architecture}
\end{figure*}

As chemical bonds exist in both graphs, $\mathcal{M}$ and $\mathcal{G}$,  acting as a bridge between them, we leverage a graph neural network that operates on $\mathcal{M}$, denoted as S-GNN in Figure~\ref{fig:architecture}(b), to learn the chemical bond's embedding $\bm {h}_{uv}$ of a protein. S-GNN updates the embedding of each chemical bond by aggregating neighborhood information in an iterative manner. Specifically, we first initialize the embeddings of chemical bonds $(u,v)$ and $(v,w)$ as $\bm {h}_{uv}^{(0)} = \bm x_{uv}$ and $\bm {h}_{vw}^{(0)} = \bm x_{vw}$, respectively, then S-GNN computes the $k$-th iteration embedding for bond $(u,v)$, denoted as $\bm {h}_{uv}^{(k)}$. The expression can be formalized as:
\begin{align}
\nonumber
    \bm{m}_{uv}^{(k)}  &= \sum_{w \in \mathcal{N}(u)}\left(\bm{h}_{uv}^{(k-1)}+\bm{h}_{uw}^{(k-1)}+ \bm {x}_{wuv}\right) \\& + \sum_{w \in \mathcal{N}(v)} \left( \bm{h}_{uv}^{(k-1)} + \bm{h}_{vw}^{(k-1)} + \bm{x}_{uvw}\right)  \\
    \bm{h}_{uv}^{(k)} & =\sigma \left(W_{S}^{(k)} \left(\bm{h}_{uv}^{(k-1)} + \bm{m}_{uv}^{(k)}\right)\right)
\end{align}
where $\mathcal{N}(u)$ and $\mathcal{N}(v)$ are the neighbors of $u$ and $v$. $\bm{m}_{uv}^{(k)}$ represents bond $(u,v)$ at iteration $k$, $W_{S}^{(k)}$ is a learnable matrix, and $\sigma$ is the activation function.

We then use another network (T-GNN, Figure~\ref{fig:architecture}(b)) to learn residue embeddings in $\mathcal{G}$. Initialized as $\bm {h}_{u}^{(0)} = \bm x_{u}$, T-GNN iterates to learn residue $u$'s embedding, $\bm{h}_{u}^{(k)}$, as:
\begin{align}
    \bm{m}_{v}^{(k)} &= \sum_{w \in \mathcal{N}(v)}\left(\bm{h}_{v}^{(k-1)} + \bm{h}_{w}^{(k-1)} + \bm{h}_{vw}^{(k-1)}\right) \\
    \bm{h}_{v}^{(k)} &= \sigma \left(W_{T}^{(k)} \left(\bm{h}_{v}^{(k-1)} + \bm{m}_{v}^{(k)}\right)\right)
\end{align}
where $\mathcal{N}(v)$ indicates the neighboring residues of $v$, $\bm{m}_{v}^{(k)}$ is denoted as the intermediate representation of residue $v$ at the $k$-th step of iteration, $\bm{h}_{vw}^{(k-1)}$ is the learned embedding of chemical bond $(v,w)$ at the $(k-1)$-th iteration in $\mathcal{M}$, and $W_{T}^{(k)}$ is a learnable matrix at the $k$-th layer.

Finally, average pooling aggregates local residue information into a global protein representation:
\begin{align}
    \bm{h}^{T} = \frac{1}{|\mathcal{V}|} \sum_{v \in \mathcal{V}} \bm{h}_{v}^{(K)}
\end{align}
where $K$ is the total iterations, and $\bm{h}^{T}$ is the protein representation, used in the pre-training module.

\subsection{Protein Representation Alignment}
Since the sequence and structure representations are learned by different encoders in different spaces, we introduce a representation alignment module to project them into a common latent space. Two mapping functions, $f_{\text{ss}}$ and $f_{\text{mvs}}$ (Figure~\ref{fig:architecture}(c)), accommodate the distinct objectives of self-supervision and multi-view supervision (Section~\ref{subsubsec:pre-training}).

Specifically, taking the mapping function $f\textsubscript{ss}$ designed for self-supervision as an example, for protein $i$ under the structure view, the transformation can be described as follows:
\begin{align}
\label{eq:fss}
    \bm z_{i}^{T} = \sigma\textsubscript{ss} (\bm W\textsubscript{ss}^{\top} \cdot \bm h_{i}^{T} + \bm b\textsubscript{ss})
\end{align}
where $\bm z_{i}^{T}$ is the projected representation of protein $i$, $\sigma\textsubscript{ss}$ is an activation function, $\bm {W}\textsubscript{ss}$ is a mapping matrix and $\bm {b}\textsubscript{ss}$ is denoted as a bias vector, respectively. Similarly, we can also obtain the transformed protein sequence representation $\bm {z}_{i}^{S}$ after the same projection transformation. 

In the same way, for the mapping function $f\textsubscript{mvs}$ tailored for multi-view supervision, the transformation for protein $i$ under the structure view can be depicted as:
\begin{align}
    \bm d_{i}^{T} = \sigma\textsubscript{mvs} (\bm W\textsubscript{mvs}^{\top} \cdot \bm h_{i}^{T} + \bm b\textsubscript{mvs})
\end{align}
where $\bm d_{i}^{T}$ is the projected representation of protein $i$. Similarly, we can also obtain the transformed sequence representation of the protein $\bm d_{i}^{S}$.

\subsection{Auxiliary Supervision for Pre-training}
\label{subsubsec:pre-training}

\subsubsection{Self-Supervision within Structure View}
\label{subsec:ss}
We leverage the NT-Xent loss~\cite{chen2020simple} to explore the self-supervision of protein structures. For each protein $\mathcal{P}$, we first extract two distinct sub-proteins $\hat{\mathcal{P}}$ and $\tilde{\mathcal{P}}$ by a specific data augmentation module, which randomly selects an amino acid residue as the center of a sphere and takes all the amino acid residues within 10.0 \AA\space to form a new protein. The objective is to maximize the mutual information between two sub-protein structure representations.

In a batch of $N$ proteins, we denote $\bm h_{i}^{\hat{T}}$ and $\bm h_{i}^{\tilde{T}}$ as the two sub-protein representations under the structure view of the $i$-th protein, using the structure encoder $g_{str}$. The mapping function $f\textsubscript{ss}$ in Figure~\ref{fig:architecture}(c) is utilized to project the representations into a lower-dimensional latent space, denoted as $\bm {z}_{i}^{\hat{T}}$ and $\bm {z}_{i}^{\Tilde{T}}$. The NT-Xent loss for the $i$-th protein is formalized as:
\begin{align}
    \mathcal{L}_{ss}^{i} = -\log \frac{\exp\left(\operatorname{sim}\left({\bm z_{i}^{\hat{T}}, \bm z_{i}^{\tilde{T}}}\right)/\tau_{s} \right)} {\sum_{k=1}^{N} \mathbf{1}\{k \neq i\} \exp \left({\operatorname{sim}{\left(\bm z_{i}^{\hat{T}}, \bm z_{k}^{\tilde{T}}\right)}}/ \tau_{s} \right)} 
\end{align}
where $\tau_{s}$ is a temperature coefficient, $\mathbf{1}\{k \neq i\} \in \{0,1\}$ is a standard indicator function that equals to 1 only when $k \neq i$ and $\operatorname{sim}(\bm u, \bm v)$ is a specified operation used to compute the cosine similarity between vectors $\bm u$ and $\bm v$.

\subsubsection{Multi-View Supervision within Sequence-Structure View}

As self-supervision only focuses on features under the structure view, we use multi-view supervision to discover the relevance between the structure and sequence views. The sub-proteins $\hat{\mathcal{P}}$ and $\tilde{\mathcal{P}}$ of protein $\mathcal{P}$ can only capture the local features of the protein, failing to represent the global features comprehensively. Hence we directly utilize the protein sequence encoder $g_{seq}$ to encode the original protein sequence $S$ without applying any stochastic data augmentations.

Given a batch of $N$ proteins, the sequence-structure pairs of the $i$-th protein can be expressed as $\{(\bm h_{i}^{S}, \bm h_{i}^{\hat{T}}),(\bm h_{i}^{S}, \bm h_{i}^{\tilde{T}})\}$.
The mapping function $f\textsubscript{mvs}$ in Figure~\ref{fig:architecture}(c) is leveraged to map each representation to the multi-view embedding space. $\bm d_{i}^{S}$, $\bm d_{i}^{\hat{T}}$ and $\bm d_{i}^{\tilde{T}}$ are the projected embeddings of the $i$-th protein sequence and sub-structures, respectively. The loss for the sub-protein $\hat{\mathcal{P}_{i}}$ of the sequence encoder is denoted as:
\begin{align}
    \mathcal{L}_{seq}^{\hat{{\mathcal{P}_{i}}}} &= - \log \frac{\exp\left(\operatorname{sim}\left(\bm d_{i}^{S}, \bm d_{i}^{\hat{T}} \right)/\tau_{m}\right)} {\sum_{k=1}^{N} \exp\left({\operatorname{sim}{\left(\bm d_{i}^{S}, \bm d_{k}^{\hat{T}}\right)}} / \tau_{m}\right)}
\end{align}
where $\tau_{m}$ is a temperature coefficient. Similarly, we can obtain the loss of sub-protein $\tilde{\mathcal{P}_{i}}$ about the sequence encoder, denoted as $\mathcal{L}_{seq}^{\tilde{\mathcal{P}_{i}}}$. The loss of protein $i$ for the sequence encoder $\mathcal{L}_{seq}^{i}$ is the average of  $\mathcal{L}_{seq}^{\hat{\mathcal{P}_{i}}}$ and $\mathcal{L}_{seq}^{\tilde{\mathcal{P}_{i}}}$. Meanwhile, we can calculate the loss for sub-protein $\hat{{\mathcal{P}_{i}}}$ of the structure encoder:
\begin{align}
    \mathcal{L}_{str}^{\hat{{\mathcal{P}_{i}}}} &= - \log \frac{\exp\left(\operatorname{sim}\left(\bm d_{i}^{\hat{T}}, \bm d_{i}^{S} / \tau_{m} \right)\right)} {\sum_{k=1}^{N} \exp \left({\operatorname{sim}{\left(\bm d_{i}^{\hat{T}}, \bm d_{k}^{S}\right)} }/ \tau_{m}\right)}
\end{align}
Following the above formulas, we can acquire the loss of sub-protein $\tilde{\mathcal{P}_{i}}$ concerning the structure encoder, denoted as $\mathcal{L}_{str}^{\tilde{\mathcal{P}_{i}}}$. The loss of protein $i$ for the structure encoder $\mathcal{L}_{str}^{i}$ is the average of  $\mathcal{L}_{str}^{\hat{{\mathcal{P}_{i}}}}$ and $\mathcal{L}_{str}^{\tilde{{\mathcal{P}_{i}}}}$ as well. Here we can obtain the loss of the $i$-th protein with multi-view supervision:
\begin{align}
    \mathcal{L}_{mvs}^{i} = \frac{1}{2} \left(\mathcal{L}_{seq}^{i} + \mathcal{L}_{str}^{i}\right)
\end{align}
So far, we have the overall loss of the pre-training phase:
\begin{align}
    \mathcal{L}_{pre} = \frac{1}{2N} \sum_{i=1}^{N} \left(\alpha \cdot \mathcal{L}_{mvs}^{i} + (1 - \alpha) \cdot \mathcal{L}_{ss}^{i} \right)
\end{align}
where the coefficient $\alpha$ is used to balance the effect of the two auxiliary supervisions. Consequently, \method \space is pre-trained via back propagation using the overall loss $\mathcal{L}_{pre}$. We utilize the pretrained structure-based encoder $g_{str}$ with an MLP as the prediction head to conduct the specific protein function prediction task in Section~\ref{exp}.

\begin{table*}
    \centering
    \tabcolsep=0.05cm
    \caption{The performance comparison with baselines on benchmark datasets. [*] denotes results taken from \protect\cite{Wang2022LMGVPAE}.}
    \label{tab:all_result}
    \begin{adjustbox}{}
        \begin{tabular}{l|cccccc|cccc}
            \toprule 
            	\multicolumn{1}{c}{} &
            \multirow{2}{*}{\textbf{Method}} &
            
            \multicolumn{1}{c}{\textbf{Pre-training Dataset}} &
            \multicolumn{4}{c|}{\textbf{F\textsubscript{max}}} &
            \multicolumn{4}{c}{\textbf{AUPR}} \\
            \cmidrule{4-8}
            \cmidrule{8-11}
            \multicolumn{1}{c}{} &
            & /\textbf{\# Parameters} &
            \multicolumn{1}{c}{\textbf{EC}} &
            \multicolumn{1}{c}{\textbf{GO-BP}} &
            \multicolumn{1}{c}{\textbf{GO-MF}} &
            \multicolumn{1}{c|}{\textbf{GO-CC}} &
            \multicolumn{1}{c}{\textbf{EC}} &
            \multicolumn{1}{c}{\textbf{GO-BP}} &
            \multicolumn{1}{c}{\textbf{GO-MF}} &
            \multicolumn{1}{c}{\textbf{GO-CC}}  \\
            
            \midrule
            
            \multirow{4}{*}{\rotatebox{90}{\textbf{Sequence}}}
            & CNN~\cite{Shanehsazzadeh2020IsTL}
            & - / 9M & 0.545& 0.244& 0.354 & 0.287 
            &0.526         &0.159          &0.351      &0.204\\
            
            & ResNet~\cite{rao2019evaluating}
            & -/ 11M & 0.605& 0.280& 0.405 & 0.304 
            &0.590         &0.205          &0.434      &0.214\\
            
            & LSTM~\cite{rao2019evaluating}
            & -/ 27M & 0.425& 0.225& 0.321 & 0.283  
            &0.414         &0.156          &0.334      &0.192\\
            
            & Transformer~\cite{rao2019evaluating}
            & - / 21M & 0.238& 0.264& 0.211 & 0.405  
            &0.218         &0.156          &0.177      &0.210\\
            
            \midrule
            \multirow{8}{*}{\rotatebox{90}{\textbf{Structure}}}

            & GCN~\cite{kipf2017semi}
            & - / 1M & 0.320& 0.252 & 0.195 & 0.329  
            &0.319        &0.136          &0.147      &0.175\\
            
            & GAT~\cite{velickovic2018graph}
            & - / 3M & 0.368& 0.284* & 0.317* & 0.385*  
            &0.320         &0.171          &0.329      &0.249\\
            
            & GVP~\cite{jing2021equivariant}
            & - /6M & 0.489& 0.326* & 0.426* & 0.420*  
            &0.482         &0.224          &0.458      &0.279\\
            
            & 3DCNN~\cite{Derevyanko2018DeepCN}
            & - / 5M & 0.077& 0.240 & 0.147 & 0.305  
            &0.029         &0.132          &0.075      &0.144\\
            
            & GraphQA~\cite{Baldassarre2020GraphQA}
            & - /2M & 0.509 & 0.308 & 0.329 & 0.413 
            &0.543         &0.199          &0.347      &0.265\\
            
            & IEConv~\cite{Hermosilla2022ContrastiveRL}
            & - / 5M & 0.735 & 0.374 &0.544 &0.444 
            &0.775         &0.273          &0.572      &0.316\\
            
            & GearNet~\cite{Zhang2022Protein}
            & - / 11M & 0.730 & 0.356 & 0.503 & 0.414
            &0.751         &0.211          &0.490      &0.276\\
            
            & GearNetEdge-Sup~\cite{Zhang2022Protein}
            & - / 21M  & 0.810 &0.403 &0.580 & 0.450 
            &0.872         &0.251          &0.570      &0.303\\
            
            \midrule
            \multirow{6}{*}{\rotatebox{90}{\textbf{Pretrained}}}
            
            & DeepFRI~\cite{Gligorijevi2021Structurebased}
            & Pfam (10M) / 709K & 0.631& 0.399& 0.465& 0.460  
            &0.547         &0.282          &0.462      &0.363\\
            
            & ESM-1b~\cite{rives2021biological}
            & UniRef50 (24M) / 650M & 0.862 & 0.452 & \underline{0.657} & 0.477 
            &0.889         &0.342          &0.639     &0.384\\
            
            & ProtBERT-BFD~\cite{elnaggar2020prottrans}
            & BFD (2.1B) / 420M & 0.838& 0.279* & 0.456* & 0.408* 
            &0.859         &0.188          &0.464      &0.234\\
            
            & LM-GVP~\cite{Wang2022LMGVPAE}
            & UniRef100 (216M) / 424M & 0.664& 0.417* & 0.545* & \bf{0.527}*  
            &0.710         &0.302          &0.580      &\underline{0.423}\\
            
            & TransFun~\cite{boadu2023combining}
            & UniRef50 (24M)) / 680M & 0.873 & 0.465  & 0.645 & 0.483  
            &\underline{0.892}         &\underline{0.348}          &\underline{0.645}      &0.391\\
            
            & GearNetEdge-Pre~\cite{Zhang2022Protein}
            & Swiss-Prot (542K) / 21M & \underline{0.878}& \underline{0.462} & 0.639 &0.465  
            &0.885         &0.283          &0.596      &0.336\\
         
            \cmidrule{2-11} 
            & \textbf{\method} & Swiss-Prot (542K) / 32M & \bf{0.891}& \bf{0.489}& \bf{0.678} & \underline{0.519} 
            &\bf{0.902}         &\bf{0.359}          &\bf{0.652}      &\bf{0.431}\\
            \bottomrule
        \end{tabular}
    \end{adjustbox}

\end{table*}

\section{Experiments}
\label{exp}
\subsection{Datasets} 
We utilize AlphaFold Swiss-Prot, a protein structure database predicted by the AlphaFold model~\cite{AlphaFold2021}, for pre-training. Meanwhile, we utilize four datasets for comparison, i.e., Enzyme Commission (EC)~\cite{Gligorijevi2021Structurebased}, Gene Ontology Molecular Function (GO-MF), Gene Ontology Cellular Component (GO-CC) and Gene Ontology Biological Process (GO-BP)~\cite{Zhang2022Protein}, all of which are used for protein function prediction. The details of benchmark datasets are described in Table~\ref{tab:dataset_detail}.
\begin{table}[!t]

    \centering
	\caption{Details on Benchmark Datasets}
	\begin{tabular}{ccccl}
	\toprule
	\textbf{Dataset}    &\# \textbf{Train}   &\# \textbf{Valid}       &\#\textbf{Test}        	&\#\textbf{Tasks} \\
	\midrule
	EC	    &15,550			&1,729		    &1,919		&538	\\
	GO-BP	&29,898			&3,322		    &3,415		&1,943    \\
    GO-MF	&29,898			&3,322	     	&3,415		&489	\\
	GO-CC   &29,898			&3,322	        &3,415		&320	\\

	\bottomrule
	\end{tabular}
	\label{tab:dataset_detail}
\end{table}

\subsection{Baselines}

To demonstrate the effectiveness of \method, we compare it against 17 baseline methods, categorized in Table~\ref{tab:all_result} as sequence-based supervised, structure-based supervised, and pre-trained methods.

\subsection{Evaluation Metrics}
We evaluate \method \space by the area under the precision-recall curve (AUPR) and protein centric maximum F-score (F\textsubscript{max})~\cite{Gligorijevi2021Structurebased}. Readers can refer to the Appendix for detailed explanations.

\subsection{Performance Comparison}
The F\textsubscript{max} and AUPR on four benchmark datasets are shown in Table~\ref{tab:all_result}, with the best results highlighted in bold and the second-best results highlighted with underlines. With respect to the F\textsubscript{max} metric, \method \space outperforms all baselines in EC, GO-BP, and GO-MF. Specifically, the F\textsubscript{max} of the experimental results have improved by $1.3\%$, $2.7\%$, and $2.1\%$ compared to the second-best results, respectively. As for the metric of AUPR,  \method \space outperforms the others in all datasets.

Compared with the sequence and structure based supervised models, \method \space outperforms all baselines, proving the pre-training phase effectively enhances the performance of \method \space on small-scale datasets. Then we evaluate the effectiveness of \method \space  with  pre-trained models. In addition to being slightly inferior to LM-GVP in the GO-CC task, \method \space outperforms other pre-training models on all metrics in other tasks. Meanwhile, we also provide the number of parameters for each pretrained model in Table~\ref{tab:all_result}. For deep learning models, having more parameters usually means higher computational demands and stronger learning capabilities. It is noteworthy that despite being only 5\% and 12\% of the parameter size of TransFun (680M) and LM-GVP (216M), respectively, SCOP (32M) still achieves outstanding performance, demonstrating its advantage of achieving high performance with relatively lower computational demands.

\begin{figure}
    \centering
    \includegraphics[scale=0.4]{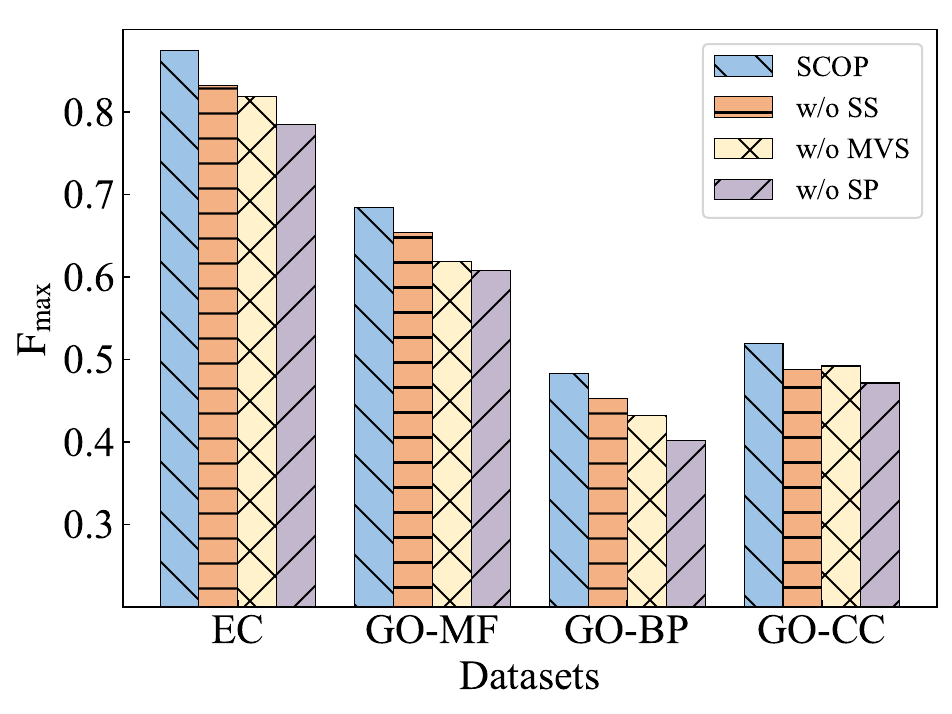}
    \caption{Results of the ablation study.}
    \label{fig:ablation}
 \end{figure}

\begin{figure*}
    \centering
    \subfigure{\includegraphics[width=0.2\linewidth]{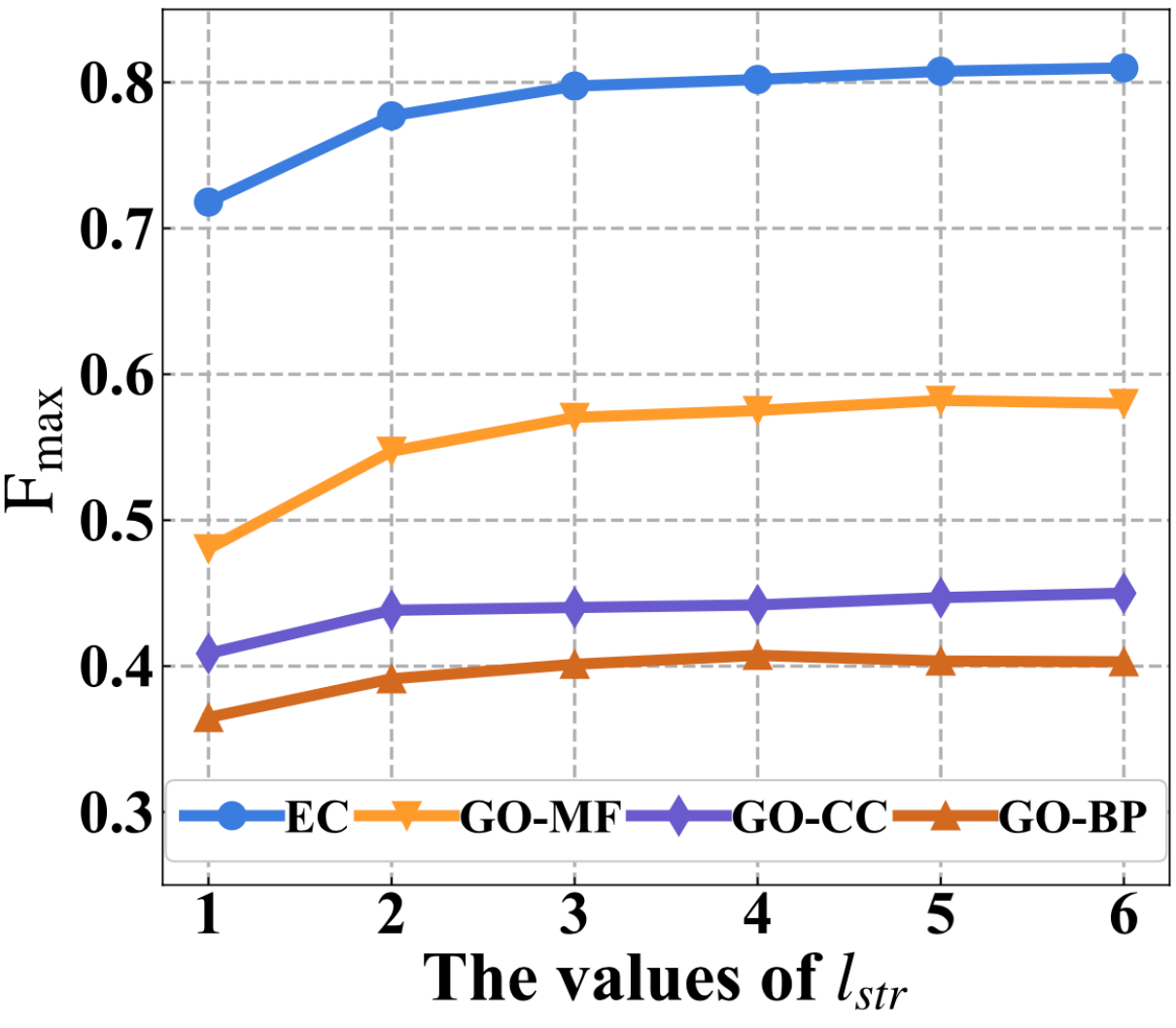}}
    \subfigure{\includegraphics[width=0.2\linewidth]{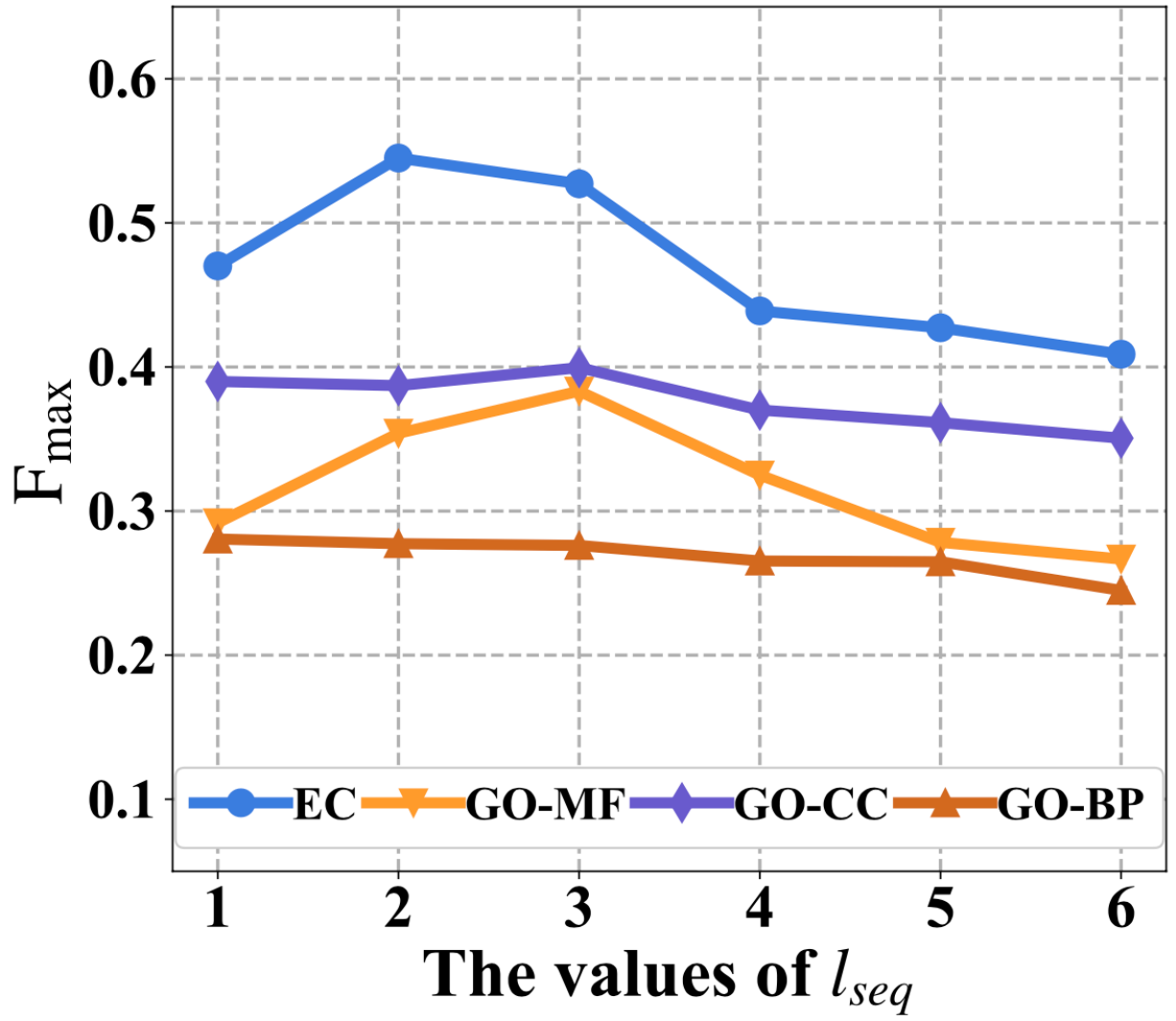}}
    \subfigure{\includegraphics[width=0.2\linewidth]{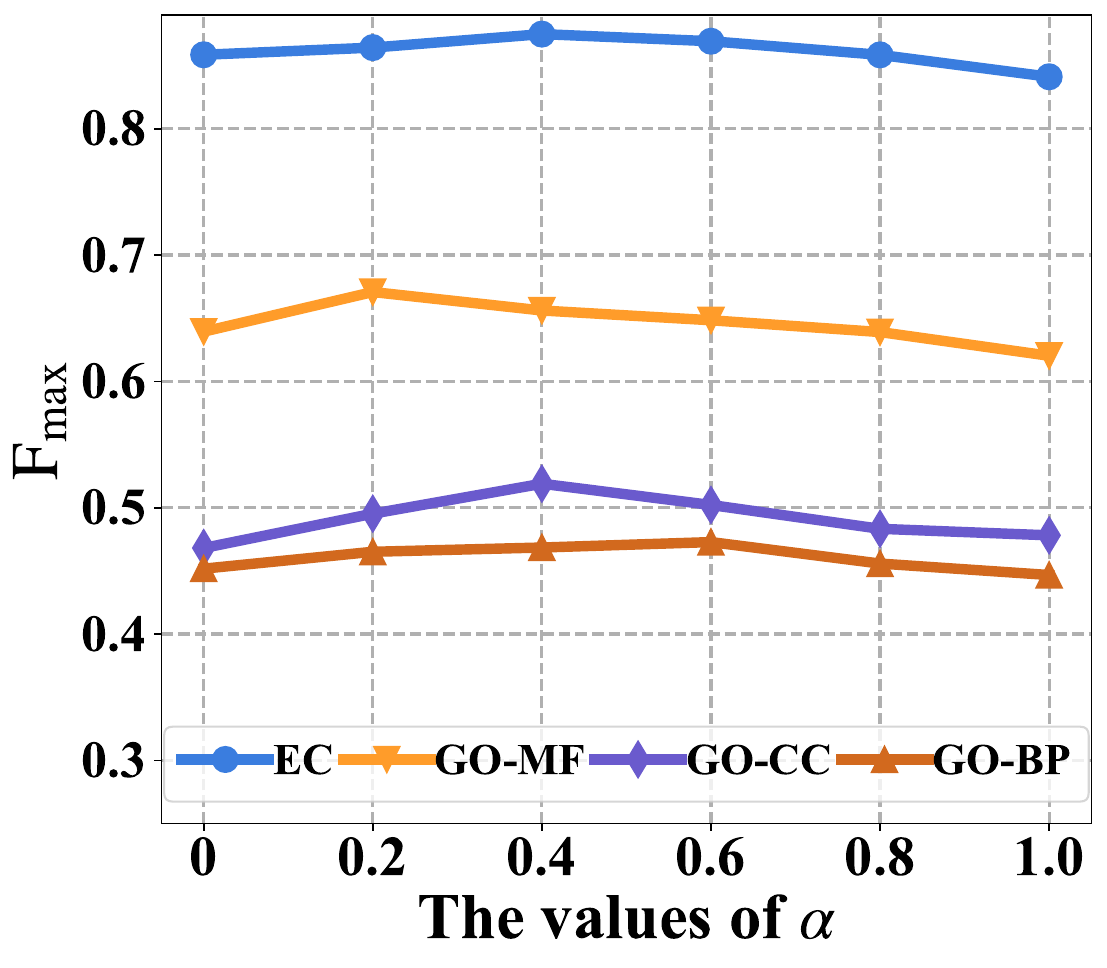}} 
    \subfigure{\includegraphics[width=0.2\linewidth]{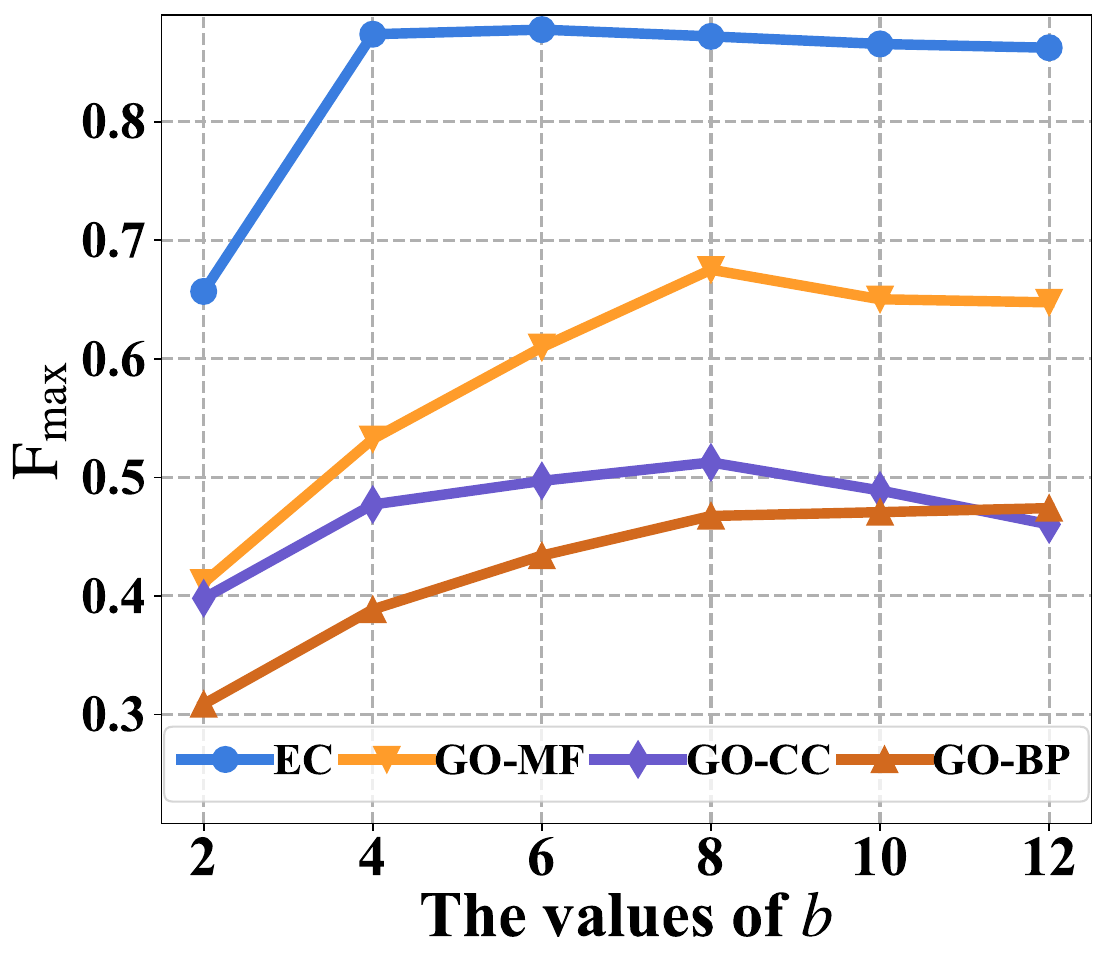}}
    \caption{The impacts of the hyperparameters.}
    \label{fig:sensitive}
\end{figure*}
 \subsection{Ablation Study}
 To gain a deeper understanding of \method \space and validate its effectiveness, we propose three variants of the model:
      (1) \textbf{w/o SS}: pretrained without self-supervision;
      (2) \textbf{w/o MVS}: pretrained without multi-view supervision;
      (3) \textbf{w/o SP}: pretrained without protein spatial information.
 Figure~\ref{fig:ablation} illustrates the performance of \method \space and its variants on four benchmark datasets. We can see that the F\textsubscript{max} of \method \space is the highest on these datasets, proving the protein spatial information and pre-training supervision tasks developed in this paper are effective.

\subsection{Parameter Sensitivity Analysis}
\label{sec:parameter}

To optimize \method's sequence-based ($l_{seq}$) and structure-based ($l_{str}$) encoder layers, we first tuned them on downstream tasks. Optimal $l_{seq}$ was 2 for EC and 3 for other tasks. For $l_{str}$, performance plateaued at 5 or greater, with further increases offering diminishing returns while increasing complexity and overfitting risk. Loss balance factor ($\alpha$) optimization (Figure~\ref{fig:sensitive}) revealed dataset-specific optima (0.4 for EC, 0.2 for GO-MF, 0.4 for GO-CC, and 0.6 for GO-BP), likely reflecting varying dataset characteristics and target-representation relationships. Structural encoder batch size ($b$) also impacted performance, with optimal values at 6 (EC), 8 (GO-MF, GO-CC), and 12 (GO-BP), though generally $b \geq$ 8 yielded satisfactory results.

\subsection{Case Study}
\label{case}

A glycoprotein dataset (about 600 positive and 900 negative samples) extracted from the RCSB PDB is used to evaluate \method's efficacy. Representations are learned using GearNetEdge-Pre, ESM-1b, and \method variations pre-trained on sequence (\method-Seq), structure (\method-Str), and combined sequence-structure (\method) views.

\begin{figure}
    \centering
    \includegraphics[width=0.8 \linewidth]{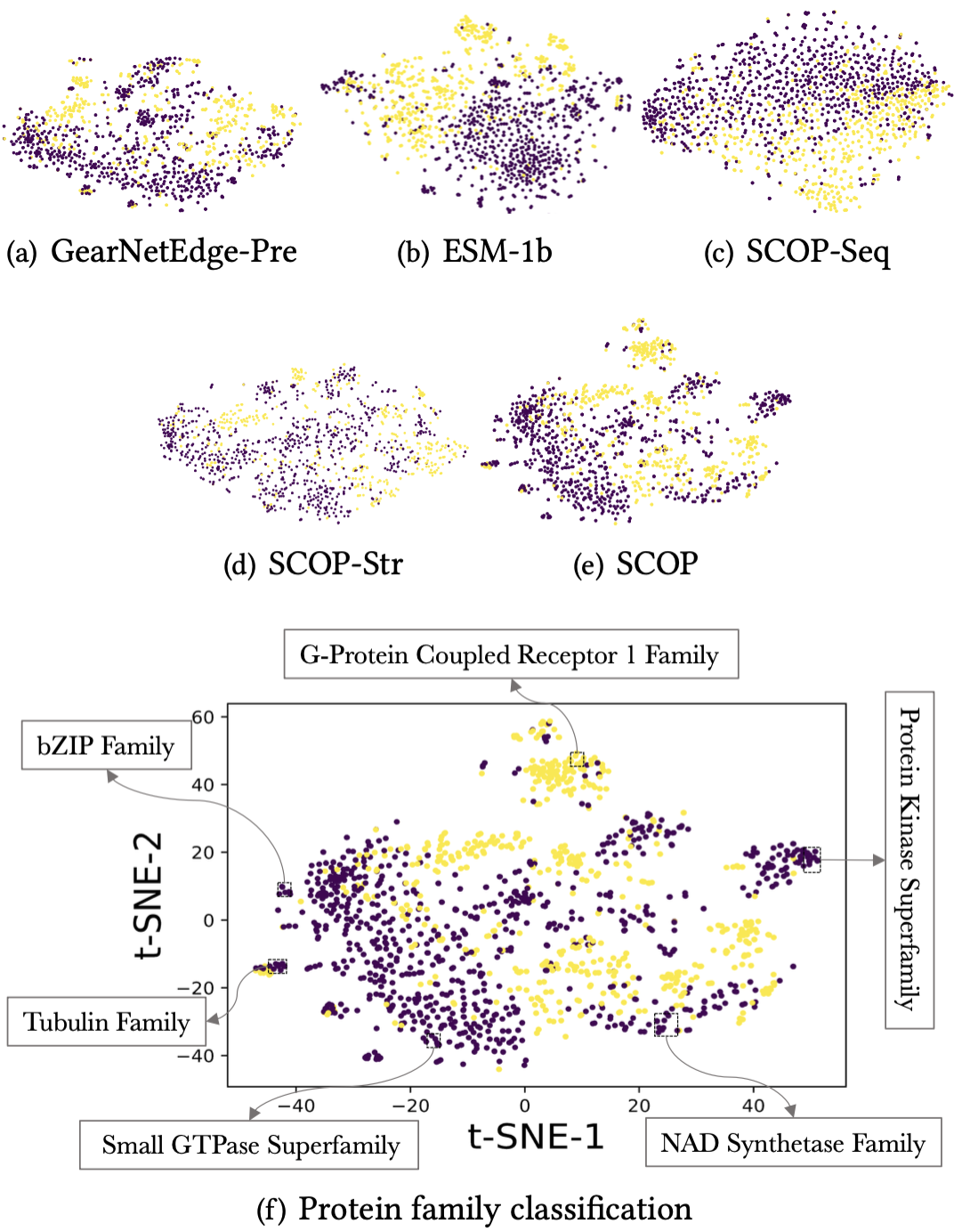}
     \caption{Visualization of protein embeddings on the proposed glycoprotein dataset. The davies-bouldin index score for (a) (b) (c) (d) (e) are 5.0164, 5.2561, 6.9586, 5.3127 and 4.2593, respectively. Proteins in the same family or superfamily tend to be clustered in the t-SNE space in (f).}
   \label{fig:case}
\end{figure}

We use t-SNE to visualize learned protein representations (Figure~\ref{fig:case}), coloring points by oligosaccharide binding ability and evaluating cluster separation with the Davies-Bouldin index (lower is better). While GearNet-Pre (5.0164) and SCOP-Str (5.3127) use only structure information, and ESM-1b (5.5261) and SCOP-Seq (6.9586) use only sequence information, \method \space leverages both, achieving the lowest index (4.2593) and suggesting more accurate representations with better glycoprotein discrimination. Furthermore, \method effectively groups proteins from the same family/superfamily (e.g., Tubulin, NAD synthetase, bZIP, and protein kinase; Figure~\ref{fig:case}(f)), demonstrating biological relevance.

\section{Conclusion}
\label{con}
In this work, we propose \method, a pre-training framework based on the contrastive supervision for protein function prediction. A novel protein encoder is introduced to integrate the protein topological and spatial information under the structure view. Moreover, the self-supervision and multi-view supervision are leveraged to exploit relevance between protein sequence and structure features during the pre-training phase so as to learn more comprehensive protein representations. Extensive experiments demonstrate that \method \space outperforms previous methods while pre-training with less data. As to future work, we intend to further explore the relationship between pre-trained protein language models and structural models to boost our model's performance.

\end{document}